\documentclass[referee]{raa}            
\usepackage{graphicx,times}             
\usepackage{subcaption}
\usepackage{graphicx}
\usepackage{natbib}
\usepackage{amssymb,amsmath}
\bibpunct{(}{)}{;}{a}{}{,}

\usepackage[pagebackref=true]{hyperref}

\begin{document}

  \title{Enhancing Galaxy Classification with U-Net Variational Autoencoders. II. JWST High Redshift Galaxy Sample}

   \volnopage{Vol.0 (20xx) No.0, 000--000}      
   \setcounter{page}{1}          

   \author{S. S. Mirzoyan 
      \inst{1,2}
   }

   \institute{Center for Cosmology and Astrophysics, Alikhanyan National Laboratory, 2 Alikhanyan Brothers str., Yerevan 0036, Armenia; {\it mserg@yerphi.am}\\
        \and
             Yerevan State University, Yerevan 0025, Armenia\\
\vs\no
   {\small Received 20xx month day; accepted 20xx month day}}

\abstract{ 
Building on our previous work, we apply a U-Net Variational Autoencoder (VAE) framework to denoise galaxy images from the James Webb Space Telescope (JWST) and enhance morphological classification. This study focuses on galaxies observed up to redshift $z \approx 8$, capturing them at early evolutionary stages where their faintness and structural complexity pose challenges for the traditional classification methods. By mitigating observational noise, our approach enables the identification of morphological features, particularly in distinguishing between disk and non-disk galaxy types. We evaluate the denoising performance using standard image quality metrics and demonstrate that the enhanced images lead to improved classification accuracy across multiple deep learning models. Our analysis of a sample of 292 galaxies up to $z = 7.69$ shows 83 galaxies classified as disk-like by the GCNN model with high confidence, of those approximately 70–80\% are of redshifts greater than 3. These findings suggest that disk-like structures can be prevalent in the early universe. The results highlight the potential of VAE-based denoising as a robust pre-processing step for analyzing high-redshift galaxy populations in ongoing astronomical surveys.
}

   \authorrunning{S. S. Mirzoyan} 
   \titlerunning{JWST High Redshift Galaxy Classification}  

   \maketitle

%
%
\section{Introduction}           
\label{sec:intro}
Machine learning (ML) has become deeply integrated into modern astronomical research, offering transformative capabilities for data analysis and interpretation ~\cite{Gh, Tian, San, Geo, Agui, Mas, At, Mirzoyan, Petrillo}. Deep learning, in particular, has transformed the landscape of galaxy morphology studies, enabling scalable and accurate classification across large astronomical datasets. Recent efforts have demonstrated the efficiency of the convolutional neural networks (CNNs) and group-equivariant architectures in capturing complex morphological features~\cite{Pandya, Hui, qian2023performance}. However, the performance of these models is often constrained by the quality of input data, particularly in the presence of noise, artifacts, and low signal-to-noise ratios—conditions that are especially prevalent in high-redshift observations.

The James Webb Space Telescope (JWST) has opened a new window into the early universe, providing unprecedented imaging depth and resolution. Recent catalogs, such as those compiled by ~\cite{Genin}, have enabled structural analysis of galaxies out to redshifts beyond $z > 10$, revealing a diverse population of compact, irregular, and evolving systems. These early-stage galaxies pose unique challenges for automated classification due to their faintness and morphological ambiguity.

In addition to morphological studies, JWST data have spurred the development of machine-learning tools for object identification across stellar and extragalactic domains. For instance, ~\cite{Crompvoets} introduced SESHAT, an XGBoost-based classifier capable of distinguishing young stellar objects, field stars, and galaxies using synthetic photometry adapted to JWST filters, achieving over 80\% recall across classes without spatial priors. Such approaches highlight the growing role of data-driven methods in maximizing JWST’s scientific return.

Spectroscopic surveys have also advanced our understanding of galaxy evolution during the reionization era. ~\cite{Meyer} present the largest spectroscopic catalog of [OIII]+HB emitters at $6.75 < z < 9.05$, constraining the [OIII] luminosity function and revealing an accelerated decline in luminosity density between $z \sim 7$ and $z \sim 8$. These findings underscore the importance of wide-area JWST slitless surveys for mapping large-scale structure and refining models of nebular line emission in the early universe.

Beyond the reionization epoch, JWST programs are probing galaxy properties at Cosmic Noon. ~\cite{Belli} describe the Blue Jay survey, a deep NIRSpec spectroscopic campaign targeting 153 galaxies at $1.7 < z < 3.5$, providing full rest-frame optical coverage and enabling robust measurements of stellar populations and gas content. This data set offers a benchmark for understanding galaxy assembly during peak star formation. Meanwhile, ~\cite{Yang} leverage JWST/PRIMER imaging to reassess the abundance of quiescent galaxies (QGs) at $z \sim 4-8$. Their analysis reveals that previous estimates were inflated by misclassified Little Red Dots, and that high-mass QGs decline sharply beyond $z > 4$, while low-mass QGs maintain nearly constant densities. These results challenge current quenching models and highlight the critical role of mid-infrared data in disentangling galaxy populations at early times.

This work we build upon our previously proposed denoising framework based on U-Net Variational Autoencoders (VAEs)~\cite{Mirzoyan2025}, which was shown to enhance classification accuracy by suppressing noise while preserving critical structural features. Here, we apply this approach to JWST NIRCam imaging data, focusing on galaxies up to redshift $z \approx 8$. Our goal is to evaluate the adaptability of the U-Net VAE model to next-generation datasets and assess its impact on the classification of disk and non-disk galaxy types in the early universe.

By integrating advanced denoising techniques with state-of-the-art classification models, this study contributes to the growing body of work aimed at improving the reliability of morphological analysis in deep-field surveys. The results offer insights into the structural evolution of galaxies at cosmic dawn and demonstrate the utility of generative models in high-redshift astrophysical research.

This paper is structured as follows: Section~\ref{sec:dataset} provides an overview of the dataset employed in our research. In Section ~\ref{sec:method}, we elaborate on the methodologies utilized for data processing and galaxy classification . Section ~\ref{sec:analysis} presents the analysis results and its significance. Lastly, Section ~\ref{sec:conclusion} concludes the paper with a summary of our findings.

\section{Dataset}
\label{sec:dataset}

This study is based on a subset of the publicly released dataset from the \textit{UNCOVER} survey \citep{Weaver}, which provides ultradeep spectroscopic observations using the \textit{James Webb Space Telescope (JWST)} NIRSpec instrument in PRISM/CLEAR mode. The full dataset includes 668 galaxies in the Abell 2744 field, spanning a redshift range of $z \sim 0.3$ to $z \sim 13$, as described in ~\cite{Prince}.

Out of the 668 galaxies, 406 have reliable spectroscopic redshift measurements. We restrict our analysis to this subset, using it to investigate galaxy morphology with the goal of classifying galaxies as either disk-like or non-disk systems. To this end, we performed a visual inspection of each galaxy, selecting only those for which structural features could be discerned despite high background noise or low surface brightness. This selection ensures that morphological classification is based on identifiable features in the imaging data, allowing for a more robust analysis of galaxy structure across cosmic time.

\begin{figure}[htbp]
    \centering
    \includegraphics[width=0.8\textwidth]{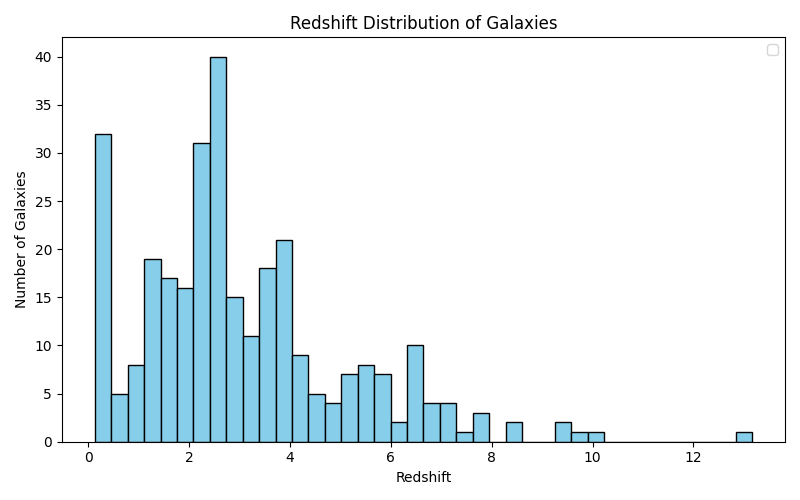}
    \caption{Histogram showing the redshift distribution of 304 galaxies. The horizontal axis represents the redshift values, while the vertical one indicates the number of galaxies in each bin.}
    \label{fig:redshift_distribution}
\end{figure}

The visual inspection resulted in a final sample of 292 galaxies, with redshifts ranging from $z = 0.172194$ to $z = 7.68681$. These galaxies were selected based on the visibility of morphological structure, enabling a meaningful classification. The redshift distribution of the final sample is summarized in Figure~\ref{fig:redshift_distribution}.

\section{Method}
\label{sec:method}

This study continues the methodology introduced in ~\cite{Mirzoyan2025}, where a U-Net Variational Autoencoder (VAE) is employed as a pre-processing step to denoise galaxy images prior to classification. The goal is to mitigate the impact of astrophysical and instrumental contamination that can obscure galaxy morphology and degrade the performance of convolutional neural networks.

We begin by selecting 1400 uncontaminated galaxy images from the EFIGI dataset ~\cite{Baillard}. Using the \texttt{PyAutoLens} package, we simulate realistic contamination by randomly inserting 1--3 non-target galaxies or stars per image, with varied light profiles, orientations, intensities, and S\'ersic indices. This ensures spatial and morphological diversity, emulating real observational conditions and dataset features, cf. e.g. \cite{DP,G}. The dataset is divided into 1000 images for training/validation and 400 for testing. To enhance generalization, each training image undergoes 10 contamination simulations, resulting in a 10,000-image augmented data set.

The images are padded to $256 \times 256 \times 3$ pixels and rescaled to the $[0, 1]$ range to match the Galaxy10 DECaLS format and stabilize training. The U-Net VAE architecture consists of an encoder with four convolutional blocks (ReLU activations, max pooling) and a decoder with transposed convolutions for upsampling. The model is trained using a combined Binary Cross-Entropy (BCE) reconstruction loss and Kullback-Leibler (KL) divergence to regularize the latent space, where BCE and KL divergence components are given by

\begin{equation}
\mathcal{L}_{\text{BCE}}(x, \hat{x}) = -\sum_{i=1}^{n} \left[ x_i \log(\hat{x}_i) + (1 - x_i) \log(1 - \hat{x}_i) \right]
\end{equation}

and 

\begin{equation}
\mathcal{L}_{\text{KL}} = -\frac{1}{2} \sum_{j=1}^{d} \left( 1 + \log(\sigma_j^2) - \mu_j^2 - \sigma_j^2 \right)
\end{equation}

respectively. Here \textit{$\mu$} and \textit{$\sigma$} are the mean and standard deviation vectors of the latent variables, and \textit{d}  is the dimensionality of the latent space. The total loss function for the Variational Autoencoder can then be expressed as:

\begin{equation}
\mathcal{L}_{\text{VAE}} = \mathcal{L}_{\text{BCE}} + \beta \mathcal{L}_{\text{KL}}
\end{equation}

where $\beta$ is a weighting factor to balance the reconstruction loss and the KL divergence.

The efficiency and performance quality of this approach are described in details in the first article of this series~\cite{Mirzoyan2025}. In contrast with the latter, where classification was performed across 10 galaxy types, this study focuses on a binary classification task: distinguishing between disk and non-disk galaxies. This simplification narrows the classification space and aligns with our scientific objective of identifying disk-like galaxies in the early universe. The JWST dataset, which includes galaxies up to redshift 13, provides a unique opportunity to explore the emergence of disk structures at cosmic dawn.

For the binary classification task, we employ Group Convolutional Neural Networks (GCNNs), which have demonstrated strong performance in prior astronomical applications. GCNNs incorporate geometric symmetries, such as rotations and reflections—into the model architecture, enhancing the robustness to spatial transformations. Following ~\cite{Pandya}, we use the \texttt{escnn} library to implement GCNNs with D16 symmetry, which includes 8 rotations and 8 reflections. This design ensures equivariance to a broad range of transformations, making the model particularly well suited for analyzing morphologically diverse, high-redshift galaxies.

\section{Analysis}
\label{sec:analysis}

Trained on the Galaxy10 DECaLS dataset, our GCNN model achieved a prediction accuracy of \textbf{98.9\%} on the training set and \textbf{98.6\%} on the test set, supported by an \textbf{F1-score of 96.5\%}. These results confirm the model’s strong generalization capability and its effectiveness in identifying disk-like structures in noisy, high-redshift galaxy images.

As far as this study involves a binary classification task—distinguishing between disk-like and non-disk-like galaxies—the model outputs a two-dimensional probability vector for each input image, representing the likelihood of the galaxy belonging to either class. Our analysis revealed that out of a total of \textbf{292} galaxies in the JWST preselected sample, the model classified \textbf{83} as disk-like. This outcome is particularly significant given the challenges associated with identifying disk features at high redshifts, where morphological signatures are often faint or distorted due to observational limitations. While a subset of these classifications may appear visually ambiguous—especially in cases where the galaxy structure is compact or irregular—the model’s predictions remain consistent with its learned latent representations and symmetry-aware architecture. This consistency suggests that the GCNN is capable of capturing subtle structural cues that may not be immediately apparent through visual inspection alone.

\begin{figure}[htbp]
    \centering
    \includegraphics[width=1.0\textwidth]{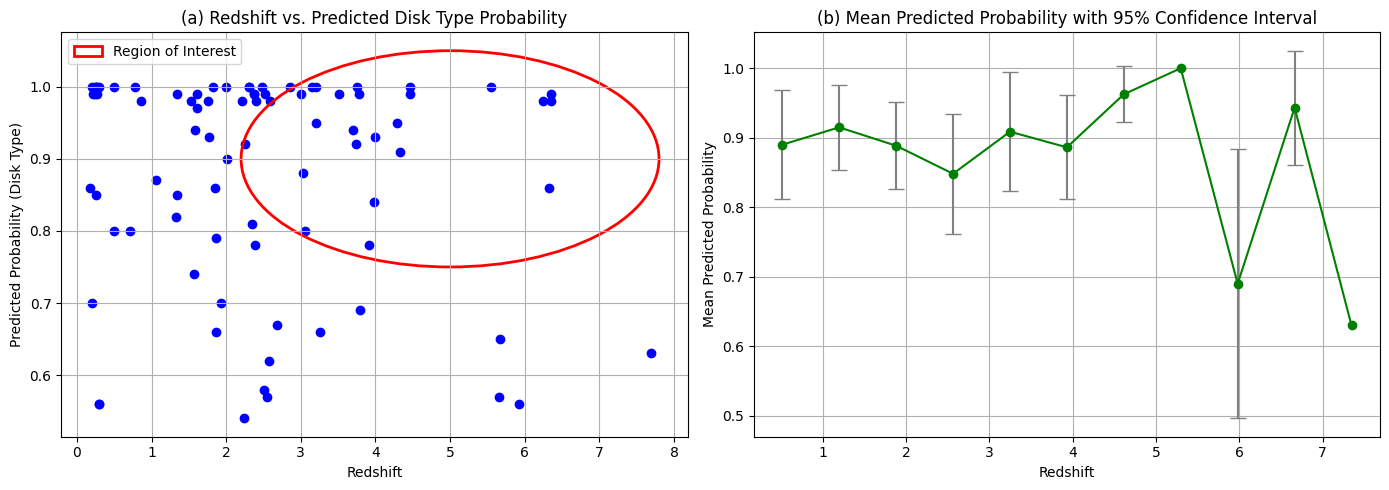}
    \caption{ \textbf{(a)} Scatter plot of spectroscopic redshift versus predicted probability of disk-like morphology for 83 galaxies classified by the GCNN model. Each point represents a galaxy identified as disk-like, with predicted probabilities exceeding 0.5. The red ellipse marks a region of interest, extending from redshifts greater than 3 and encompassing relatively higher predicted probabilities, indicating a concentration of high-confidence disk-like galaxies at intermediate to high redshifts.
    \textbf{(b)} Mean predicted probability of disk-like morphology as a function of redshift, binned into 11 intervals. Error bars represent the 95\% confidence interval of the mean in each bin, illustrating the variation and reliability of model predictions across redshift.}
    \label{fig:redshift_prob_confidence}
\end{figure}

Figure~\ref{fig:redshift_prob_confidence} \textbf{(a)} is the illustration of scatter plot of the relationship between spectroscopic redshift and the predicted probability of being disk-like for 83 galaxies identified by the GCNN model. Each blue point represents a galaxy classified as disk-like, with predicted probabilities exceeding 0.5, consistent with the binary classification threshold.

The red ellipse marks a region of interest, extending from redshifts greater than 3 and encompassing relatively higher predicted probabilities. This region highlights where a notable concentration of high-confidence disk-like galaxies may be found, particularly at intermediate to high redshifts.

To further investigate the relationship between redshift and the model's confidence in disk-like morphology classification, we computed the mean predicted probability within 11 redshift bins. As shown in Figure~\ref{fig:redshift_prob_confidence} \textbf{(b)}, each point represents the average probability in a bin, with vertical error bars indicating the confidence interval 95\% of the mean. This analysis reveals how the GCNN model's predictions vary across cosmic time, highlighting regions of stable or elevated confidence in disk-like structures, particularly at intermediate redshifts.

To facilitate further investigation and validation, the complete list of galaxy IDs identified as disk-like is provided in Table~\ref{tab:disk_galaxies}. This table serves as a reference for astronomers interested in conducting follow-up studies, including spectroscopic analysis, kinematic modeling, or deeper morphological assessments using complementary datasets.

\begin{table}[]
\centering
\begin{tabular}{|c|c|c|c|c|}
\hline
ID & $z_{\mathrm{spec}}$ & RA & Dec & Predicted Probability \\
\hline
11432 & 3.204713 & 3.607424725 & -30.40646593 & 1.0 \\
11714 & 1.5736 & 3.588383257 & -30.40524863 & 0.74 \\
12065 & 5.544374 & 3.570058411 & -30.4036885 & 1.0 \\
13177 & 0.211643 & 3.594568212 & -30.40143574 & 0.99 \\
14267 & 0.497962 & 3.590319031 & -30.40037353 & 0.8 \\
14392 & 0.290691 & 3.57905448 & -30.4000657 & 1.0 \\
14399 & 3.989351 & 3.576389822 & -30.40252961 & 0.93 \\
14880 & 1.82895 & 3.586331526 & -30.40154576 & 1.0 \\
15350 & 1.345048 & 3.596723494 & -30.39680982 & 0.85 \\
15989 & 3.696915 & 3.571727083 & -30.39561626 & 0.94 \\
17420 & 3.914086 & 3.566452615 & -30.39241038 & 0.78 \\
18023 & 2.403845 & 3.566944484 & -30.39134312 & 0.98 \\
18407 & 3.978884 & 3.562749053 & -30.39094896 & 0.84 \\
18471 & 2.211075 & 3.586734331 & -30.39077692 & 0.98 \\
18708 & 0.293699 & 3.580934582 & -30.39078092 & 0.56 \\
18924 & 7.68681 & 3.581044323 & -30.38956069 & 0.63 \\
19007 & 3.203452 & 3.555707506 & -30.39034508 & 0.95 \\
19371 & 0.27453 & 3.588066477 & -30.38866119 & 0.99 \\
19425 & 3.787967 & 3.570316257 & -30.38856369 & 0.69 \\
19958 & 2.50856 & 3.57211999 & -30.38750435 & 0.58 \\
20129 & 3.736875 & 3.564213901 & -30.38725451 & 0.92 \\
21111 & 3.056035 & 3.58249888 & -30.3854613 & 0.8 \\
22043 & 0.297067 & 3.569327118 & -30.38423508 & 0.56 \\
22084 & 3.15165 & 3.632125846 & -30.38344062 & 1.0 \\
22755 & 1.610406 & 3.608297992 & -30.38230371 & 0.99 \\
22876 & 2.48364 & 3.636093317 & -30.38190207 & 1.0 \\
22991 & 2.236867 & 3.559313307 & -30.38194215 & 0.54 \\
23889 & 1.750486 & 3.574411737 & -30.38014743 & 0.98 \\
23910 & 2.553759 & 3.54384577 & -30.38048629 & 0.57 \\
24714 & 6.352246 & 3.628748131 & -30.37890927 & 0.99 \\
24817 & 5.662414 & 3.620128014 & -30.37873014 & 0.57 \\
25197 & 0.500661 & 3.57579528 & -30.3781589 & 1.0 \\
25447 & 6.247132 & 3.569435091 & -30.37788729 & 0.98 \\
25558 & 2.343793 & 3.545584927 & -30.37779016 & 0.81 \\
25931 & 4.288877 & 3.575532423 & -30.37865006 & 0.95 \\
26882 & 1.930889 & 3.59938196 & -30.37594427 & 0.7 \\
28284 & 4.329758 & 3.623366643 & -30.37478743 & 0.91 \\
28399 & 0.265133 & 3.552684962 & -30.3818944 & 1.0 \\
28430 & 0.242316 & 3.55670675 & -30.37861348 & 0.99 \\
29244 & 2.515936 & 3.625254921 & -30.37347943 & 0.99 \\
30673 & 0.717342 & 3.606978005 & -30.37076671 & 0.8 \\
30790 & 0.855729 & 3.625604263 & -30.37083042 & 0.98 \\
31232 & 2.367282 & 3.545605817 & -30.37405148 & 0.99 \\
31927 & 2.368629 & 3.547034692 & -30.36945236 & 0.99 \\
33157 & 0.778385 & 3.54078227 & -30.36843673 & 1.0 \\
33744 & 0.202951 & 3.607382682 & -30.3658209 & 0.7 \\
34310 & 2.585887 & 3.583190975 & -30.36618681 & 0.98 \\
\hline
\end{tabular}
\end{table}

\begin{table}[ht]
\centering
\begin{tabular}{|c|c|c|c|c|}
\hline
ID & $z_{\mathrm{spec}}$ & RA & Dec & Predicted Probability \\
\hline
35436 & 2.303907 & 3.534222639 & -30.36552596 & 1.0 \\
35849 & 1.997497 & 3.609666989 & -30.36202553 & 1.0 \\
36689 & 1.859669 & 3.54777932 & -30.36282393 & 0.79 \\
36691 & 2.013727 & 3.546526021 & -30.36106014 & 0.9 \\
37187 & 0.23971 & 3.531233475 & -30.36078649 & 1.0 \\
37844 & 0.254927 & 3.570830536 & -30.35860582 & 0.85 \\
38987 & 1.062268 & 3.609138843 & -30.35684001 & 0.87 \\
41216 & 0.254927 & 3.533301414 & -30.35498935 & 1.0 \\
41851 & 0.208135 & 3.578862544 & -30.35132378 & 1.0 \\
42041 & 2.246267 & 3.581923148 & -30.35180436 & 0.92 \\
42203 & 1.324502 & 3.563828463 & -30.36763464 & 0.82 \\
42280 & 2.996847 & 3.562646955 & -30.36874216 & 0.99 \\
42328 & 5.666573 & 3.581659016 & -30.3506007 & 0.65 \\
42360 & 1.529461 & 3.571637385 & -30.35075373 & 0.98 \\
43078 & 1.347864 & 3.568650773 & -30.34918531 & 0.99 \\
43239 & 0.172194 & 3.576038833 & -30.34966523 & 0.86 \\
44387 & 2.580053 & 3.563346091 & -30.34651683 & 0.62 \\
45889 & 1.853019 & 3.567129886 & -30.34342503 & 0.86 \\
46937 & 4.466322 & 3.57934047 & -30.34199404 & 1.0 \\
46973 & 3.024923 & 3.572150878 & -30.34152605 & 0.88 \\
47040 & 4.457583 & 3.580625484 & -30.34125755 & 0.99 \\
47875 & 1.86654 & 3.585372873 & -30.3409346 & 0.66 \\
48006 & 1.577391 & 3.521258001 & -30.34757079 & 0.94 \\
48070 & 3.255778 & 3.57727173 & -30.33835157 & 0.66 \\
48208 & 1.611451 & 3.5638014 & -30.33881164 & 0.97 \\
50379 & 1.775153 & 3.559976444 & -30.33209503 & 0.93 \\
51076 & 5.929627 & 3.5536951 & -30.33005479 & 0.56 \\
51398 & 3.785574 & 3.575060222 & -30.33130673 & 0.99 \\
60018 & 2.311847 & 3.574879169 & -30.35606972 & 1.0 \\
60046 & 3.515004 & 3.556483332 & -30.37639 & 0.99 \\
60061 & 3.749341 & 3.551117917 & -30.37475583 & 1.0 \\
60076 & 2.385176 & 3.552054167 & -30.37716139 & 0.78 \\
6291 & 2.675929 & 3.605603035 & -30.41812688 & 0.67 \\
7309 & 6.346668 & 3.611317803 & -30.41455737 & 0.98 \\
8943 & 6.325826 & 3.614087404 & -30.41044796 & 0.86 \\
9457 & 2.850896 & 3.578220177 & -30.40850853 & 1.0 \\
\hline
\end{tabular}
\caption{Disk-like galaxies identified by the GCNN model, including spectroscopic redshift ($z_{\mathrm{spec}}$), right ascension (RA), declination (Dec), and predicted probability of disk-like morphology.}
\label{tab:disk_galaxies}
\end{table}

\section{Conclusions}
\label{sec:conclusion}
In this study, we presented a robust framework for enhancing morphological classification of high-redshift galaxies using a U-Net Variational Autoencoder (VAE) for image denoising, followed by a Group Convolutional Neural Network (GCNN) for binary classification of disk-like versus non-disk-like systems. By applying this pipeline to JWST NIRCam imaging data from the UNCOVER survey, we demonstrated that denoising significantly improves the visibility of structural features, enabling more reliable classification even in the presence of observational noise and low surface brightness. Our study focuses on a binary classification task: distinguishing between disk and non-disk galaxies.

Our analysis focused on a visually curated sample of 292 galaxies with reliable spectroscopic redshifts, spanning $z = 0.17$ to $z = 7.69$. Out of these, 83 galaxies were classified as disk-like by the GCNN model. Notably, a substantial fraction of these—approximately 70–80\%—are located at redshifts greater than 3 and were predicted with high confidence (probabilities well above 0.7). This finding is particularly compelling, as it suggests that disk-like structures may be more prevalent in the early universe than previously assumed, and that their morphological signatures can be effectively recovered through advanced denoising and symmetry-aware classification techniques.

These results underscore the potential of combining generative models with equivariant deep learning architectures to probe galaxy evolution at cosmic dawn. The ability to identify disk-like galaxies at such early epochs opens new avenues for studying the formation and stability of galactic disks under extreme physical conditions. Future work will extend this approach to larger samples and incorporate kinematic data to further validate the morphological classifications presented here.

\begin{acknowledgements}
{{We are thankful to the referee for helpful suggestions and comments. We acknowledge the use of the EFIGI and Galaxy10 DECaLS datasets, which enabled the development and validation of this research. Our sincere thanks are due to Artur Hakobyan for valuable insights and support in guiding the dataset selection and contributing in key discussions.}}
\end{acknowledgements}


\begin{thebibliography}{99}


\bibitem[Aguilar-Arguello et al. 2025]{Agui} Aguilar-Arguello, G. et al., 2025, Morphological Classification of Galaxies Through Structural and Star Formation Parameters Using Machine Learning, arXiv:2501.06340

\bibitem[Atemkeng et al. 2025]{At} Atemkeng, M.T. et al., 2025, A benchmark analysis of saliency-based explainable deep learning methods for the morphological classification of radio galaxies, arXiv:2502.17207 

\bibitem[Baillard et al. 2011]{Baillard} Baillard, A. et al., 2011, The EFIGI catalogue of 4458 nearby galaxies with detailed morphology, Astronomy \& Astrophysics, 532, A74

\bibitem[Belli et al. 2025]{Belli} Belli, S. et al., 2025, The Blue Jay Survey: Deep JWST Spectroscopy for a Representative Sample of Galaxies at Cosmic Noon, arXiv:2510.11775

\bibitem[Crompvoets et al. 2025]{Crompvoets} Crompvoets, B. L. et al., 2025, Object Classification from JWST Catalogs, arXiv:2510.07747

\bibitem[De Paolis et al. 2014]{DP} De Paolis, F. et al, 2014, Planck confirmation of the disk and halo rotation of M 31, Astronomy \& Astrophysics, 565, L3

\bibitem[Genin et al. 2025]{Genin} Genin, A. et al., 2025, DAWN JWST Archive: Morphology from profile fitting of over 340,000 galaxies in major JWST fields, Morphology evolution with redshift and galaxy type, arXiv:2505.21622

\bibitem[Georgiou et al. 2025]{Geo} Georgiou, C. et al., 2025, Intrinsic galaxy alignments in the KiDS-1000 bright sample: dependence on colour, luminosity, morphology and galaxy scale,  arXiv:2502.09452

\bibitem[Ghaderi et al. 2025]{Gh} Ghaderi, H. et al., 2025, Galaxy Morphological Classification with Zernike Moments and Machine Learning Approaches, arXiv:2501.09816

\bibitem[Gurzadyan 1999]{G} Gurzadyan V.G., 1999, Kolmogorov complexity as a descriptor of cosmic microwave background maps, Europhysics Letters, 46, 114

\bibitem[Hui et al. 2022]{Hui} Hui, W. et al., 2022, Galaxy Morphology Classification with DenseNet, Journal of Physics: Conference Series, Volume 2402, Issue 1, id.012009, 11 pp.

\bibitem[Masters 2025]{Mas} Masters, K., 2025, Morphological Classification of Galaxies, arXiv:2502.09610

\bibitem[Meyer et al. 2025]{Meyer} Meyer, R. A. et al., 2025, JWST COSMOS-3D: Spectroscopic Census and Luminosity Function of [O III] Emitters at $6.75<z<9.05$ in COSMOS, arXiv:2510.11373

\bibitem[Mirzoyan et al. 2019]{Mirzoyan} Mirzoyan, S. S. et al., 2019, Machine learning and Kolmogorov analysis to reveal gravitational lenses, MNRAS (Letters), Volume 489, Issue 1, p.L32-L36

\bibitem[Mirzoyan 2025]{Mirzoyan2025} Mirzoyan, S. S., 2025, Enhancing Galaxy Classification with U-Net Variational Autoencoders for Image Denoising, Research in Astronomy and Astrophysics, Volume 25, Issue 9, id.095006, 9 pp.

\bibitem[Pandya et al. 2023]{Pandya} Pandya, S. et al., 2023, E(2) Equivariant Neural Networks for Robust Galaxy Morphology Classification, arXiv:2311.01500

\bibitem[Petrillo et al 2017]{Petrillo} Petrillo, C. E.  et al., 2017, Finding strong gravitational lenses in the Kilo Degree Survey with Convolutional Neural Networks,  MNRAS, Volume 472, Issue 1, p.1129-1150.

\bibitem[Prince et al. 2025]{Prince} Prince, S. H. et al., 2025, The UNCOVER Survey: First Release of Ultradeep JWST/NIRSpec PRISM Spectra for $~$700 Galaxies from z $~$ 0.3–13 in A2744, The Astrophysical Journal, Volume 982, Issue 1, id.51, 16 pp.

\bibitem[Sanders et al. 2025]{San} Sanders, J.S. et al., 2025, The SRG/eROSITA all-sky survey: The morphologies of clusters of galaxies I: A catalogue of morphological parameters, arXiv:2502.02239

\bibitem[Tian et al. 2025]{Tian} Tian, C. et al., 2025,  Automatic Machine Learning Framework to Study Morphological Parameters of AGN Host Galaxies within z<1.4 in the Hyper Supreme-Cam Wide Survey, arXiv:2501.15739

\bibitem[Weaver et al. 2024]{Weaver} Weaver, J. R. et al. , 2024, The UNCOVER Survey: A First-look HST + JWST Catalog of 60,000 Galaxies near A2744 and beyond, The Astrophysical Journal Supplement Series, Volume 270, Issue 1, id.7, 23 pp.

\bibitem[Yang et al. 2025]{Yang} Yang, T., 2025, A census of quiescent galaxies across  $0.5 < z < 8$ with JWST/MIRI: Mass-dependent number density evolution of quiescent galaxies in the early Universe, arXiv:2510.12235

\bibitem[Yumeng 2023]{qian2023performance} Yumeng, Q., 2023, Performance comparison among VGG16, InceptionV3, and resnet on galaxy morphology classification, Journal of Physics: Conference Series, Volume 2580, Issue 1, id.012009, 7 pp.

\end{thebibliography}

\label{lastpage}

\end{document}